\documentclass[useAMS,usenatbib]{mn2e}
\usepackage[american]{babel}
\usepackage{mathrsfs,eucal,graphicx,color,amsmath,amssymb,amsfonts}
\usepackage{footnote}
\usepackage[justification=centering]{caption}
\usepackage{color}
\title[ Radio Galaxy populations and the multi-tracer technique]{ Radio Galaxy populations and the multi-tracer technique: pushing the limits on primordial non-Gaussianity}
\author[L. D. Ferramacho et al.]{L. D. Ferramacho$^{1,2}$\thanks{E-mail:
luis.ferramacho@ist.utl.pt},  M. G. Santos$^{2,1,3}$, M. J. Jarvis$^{4,2}$ and S. Camera$^{1,2}$ \\
$^{1}$CENTRA, Instituto Superior T\'ecnico, Universidade de Lisboa, Av. Rovisco Pais 1, 1049-001, Lisboa, Portugal.\\
$^{2}$Department of Physics, University of Western Cape, Cape Town 7535, South Africa.\\
$^{3}$SKA SA, 3rd Floor, The Park, Park Road, Pinelands, 7405, South Africa\\
$^{4}$Astrophysics, Department of Physics, Keble Road, Oxford OX1 3RH, UK.}

\begin{document}



\maketitle

\label{firstpage}

\begin{abstract}

We explore the use of different radio galaxy populations as tracers of different mass halos and therefore, with different bias properties, to constrain primordial non-Gaussianity of the local type.
We perform a Fisher matrix analysis based on the predicted auto and cross angular power spectra of these populations, using simulated redshift distributions as a function of detection flux and the evolution of the bias for the different galaxy types (Star forming galaxies, Starburst galaxies, Radio-Quiet Quasars, FRI and FRII AGN galaxies).
We show that such a multi-tracer analysis greatly improves the information on non-Gaussianity by drastically reducing the cosmic variance contribution to the overall error budget. By using this method applied to future surveys, we predict a constraint of $\sigma_{f_{nl}}=3.6$ on the local non-Gaussian parameter for a galaxy detection flux limit of 10$ \mu$Jy and $\sigma_{f_{nl}}=2.2$ for 1 $\mu$Jy. We show that this significantly improves on the constraints obtained when using the whole undifferentiated populations ($\sigma_{f_{nl}}=48$ for 10 $\mu$Jy and $\sigma_{f_{nl}}=12$ for 1 $\mu$Jy). We conclude that continuum radio surveys alone have the potential to constrain primordial non-Gaussianity to an accuracy at least a factor of two better than the present constraints obtained with Planck data on the CMB bispectrum, opening a window to obtain $\sigma_{f_{nl}}\sim1$ with the Square Kilometer Array.
\end{abstract}


\section{Introduction}

In the current standard cosmological model, the large scale structures observed in the Universe originated from small fluctuations present in the matter density field arising after the inflationary phase shortly after the Big Bang. Although slow-roll inflation models predict this random field to be essentially Gaussian \citep{Maldacena, Acquaviva}, other evolutionary models after inflation predict a non-vanishing non-Gaussian component in the primordial matter density field \citep{Verde, Liguori}. The detection of non-Gaussianity could open a new window in the knowledge of early Universe physics.

The most widely used method so far to constrain primordial non-Gaussianity is to measure the bi-spectrum in the Cosmic Microwave Background (CMB) temperature anisotropy maps. Recently, this method was applied to Planck data \citep{Planck2} to provide the best measurement of local non-Gaussianity up to date ($f_{nl}=2.7 \pm 5.8$)\footnote{Note that in our definition of $f_{nl}$, with the growth factor normalized to unity today, this value should be multiplied by a factor of $\approx 1.3$ \citep[see e.g.][]{Dalal,Afshordi}.}. A complementary way to access non-Gaussianity is to measure its impact on Large Scale Structure (LSS) at lower redshifts \citep{Dalal,b2, carbone} which affects the bias of dark matter tracers. Large galaxy surveys have been the probe of choice to constrain the clustering properties of dark matter on large scales and measure this non-Gaussian effect \citep{Xia2010, Xia2011, Bernardis}. Given the fact that this non-Gaussian signal is specially sensitive on large scales, other type of surveys, based on intensity mapping techniques, have been suggested to go after this effect \citep{Joudaki, Camera2013}.

Although observing large scales has the advantage of probing the regime where the non-Gaussian effect on the bias of the dark matter tracers is stronger, it has the problem that it is nonetheless limited by cosmic variance, e.g., the lack of enough independent measurements for the scales we are trying to probe, given the limited size of the volume that is observed. \citet{Seljak} proposed a way to get around this cosmic variance limitation by using different biased tracers of the underlying dark matter distribution. With at least two tracers with different bias, we can make a measurement of the ratio of these two biases that is only limited by shot noise and hence beats cosmic variance. This is specially sensitive when the bias of one object is much larger than the other. In order to understand this, let us assume that we measure two density maps at a given redshift, one for a biased tracer and another for the dark matter itself. Due to cosmic variance, there will be several power spectra (e.g. several cosmologies) that are consistent with the dark matter map. However, the ratio of the two maps should give a direct measurement of the bias, with an uncertainty just given by the shot noise of the tracer.

In this paper we propose to use the mass of the haloes hosting the dark matter tracers as an extra source of information in order to constrain the bias. This will have 3 advantages: 1) it will allow us to select objects with large bias factors (without any mixture with low mass objects) making them more sensitive to the non-Gaussian effect; 2) it will allow us to compare directly the bias of different tracers in order to cancel cosmic variance and 3) it will allow us to have a more physical description of the bias parameters used in the analysis. Although this idea can be applied to different surveys, it can be particularly relevant to radio galaxy surveys, given that they usually lack redshift information but on the other hand should provide a more direct relation between bias and mass. In here we concentrate on surveys that can be achieved with the SKA (http://www.skatelescope.org) and its pathfinders.
                  
The outline for this work is the following: In Sec. 2 we describe how to model non-Gaussian features in LSS bias, and their mass dependence. In Sec. 3 we present the forecasts for future radio surveys, focusing on the improvement of the constraints on $f_{nl}$ when using tracers of different masses with standard angular power spectrum measurements. In Sec. 4, we show how to improve these constraints by eliminating cosmic variance using multi tracer bias. Finally, in Sec. 5 we discuss the results and present our conclusions. 

\section[]{Primordial non-Gaussianity in LSS clustering}
\label{section2}
\subsection{The effect on bias}

In most standard inflationary scenarios, non-Gaussianity in the primordial fluctuation field is characterized by a local feature of the Bardeen gauge invariant potential {$\Phi$}, which is expressed as:
\begin{equation}
\Phi=\phi+f_{nl}(\phi^2-\langle \phi \rangle^2)
\label{eq:1}
\end{equation}
where $\phi$ is a random Gaussian field and $f_{nl}$ is a constant that defines the amplitude of non-Gaussianity. On sub-horizon scales, the Bardeen potential simply becomes the opposite of the gravitational potential. 
In terms of large scale structure clustering in the Universe, the main consequence of such deviation from a Gaussian field is to increase power on large scales in 2-point statistics such as the tridimensional power spectrum of clustered objects. A scale dependent correction appears in the total halo bias: \citep{Dalal,b2}
\begin{equation}
b_h(M,z)=b_L(M,z)+f_{nl}\delta_c \left[(b_L(M,z)-1\right]\frac{3 \Omega_m H_0^2}{c^2 k^2 T(k) D(z)}
\label{eq:2}
\end{equation}    
The total halo bias thus depends on the non-Gaussianity amplitude $f_{nl}$, the critical overdensity for spherical collapse $\delta_c$, at redshift z=0, as well as the linear transfer ($T(k)$) and growth ($D(z)$) functions (with $D(0)=1$). This correction is placed upon the usual Gaussian linear bias, which assumes the expression:
\begin{equation}
b_L(M,z)=1+\frac{q \nu -1}{\delta_{c}(z)}+\frac{1}{\delta_c(z)}\frac{2p}{1+(q\nu)^p}
\label{eq:3}
\end{equation}     
where $\nu=\delta_c^2(z)/\sigma_0^2(M)$ and $\delta_c(z)= D(z) \delta_c$. This model was first proposed in \cite{MoWhite} and \cite{SethTormen} obtained the values for the parameters which better fit numerical simulations of dark matter collapse and galaxy formation, finding $p=0.3$ and $q=0.75$.

\subsection{Mass dependence of non-Gaussian effects in the power spectra}

For a given tracer with bias $b_h(z)$, the total 3D halo power spectrum is then simply $P^{3D}_h=b_h^2 P_{\delta}$ (on large scales), where $P_{\delta}$ is the underlying dark matter power spectrum. If no redshift information is available to compute the 3D power spectrum, the adequate estimator to use is the angular power spectrum, where the distribution of halos along the line of sight is projected over the 2D field of view. If several tracers are considered, the total statistical information is described by the auto and cross correlation power spectra. The result consists in a set of multipole values $C_l^{i,j}$ and the full computation reads \citep{Huterer}:
\begin{equation}
C_l^{i,j}=\frac{2}{\pi}\int_{k_{min}}^{k_{max}} k^2 P_\delta(k) W^i_l(k)W^j_l(k)dk
\label{eq:4}
\end{equation}  
where $W^i_l$ is a window function which accounts for the clustering properties of a given biased tracer $i$ and the angular geometry for a given multipole $l$:
\begin{equation}
W^i_l=\int \frac{dn}{dz}^iD(z)b_h^i(z)j_l(k r)dz\;.
\label{eq:5}
\end{equation}
Here $dn/dz$ is the angular redshift distribution of sources normalized to unity, $r$ is the comoving radial distance to redshift $z$ and $j_l$ are the spherical Bessel functions of order $l$. 
\begin{figure*}
  \hspace{-9pt}
  \begin{minipage}[b]{8.6cm}
    \includegraphics[width=1\textwidth]{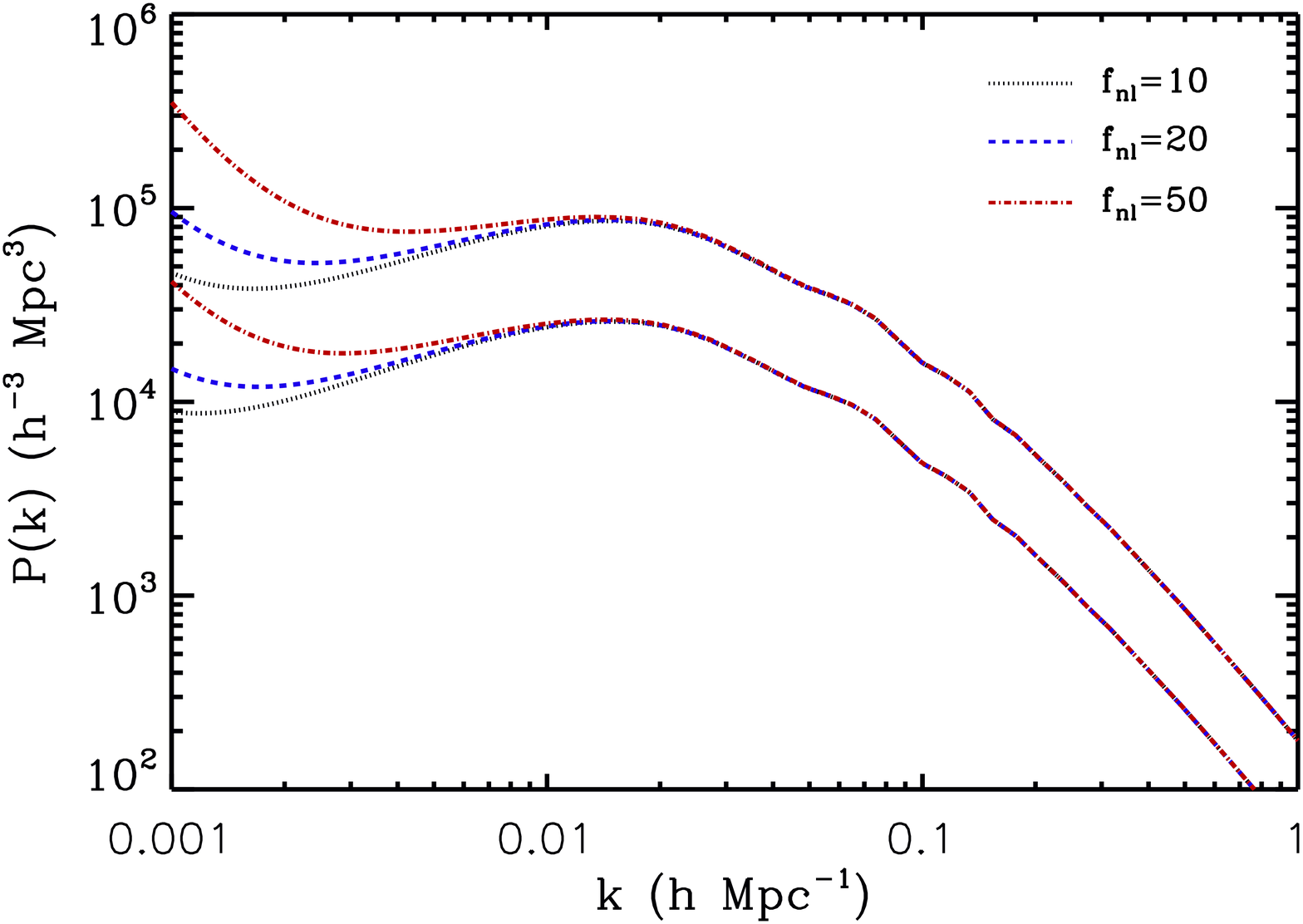}  
  \end{minipage}
  \begin{minipage}[b]{8.6cm}
    \includegraphics[width=1\textwidth]{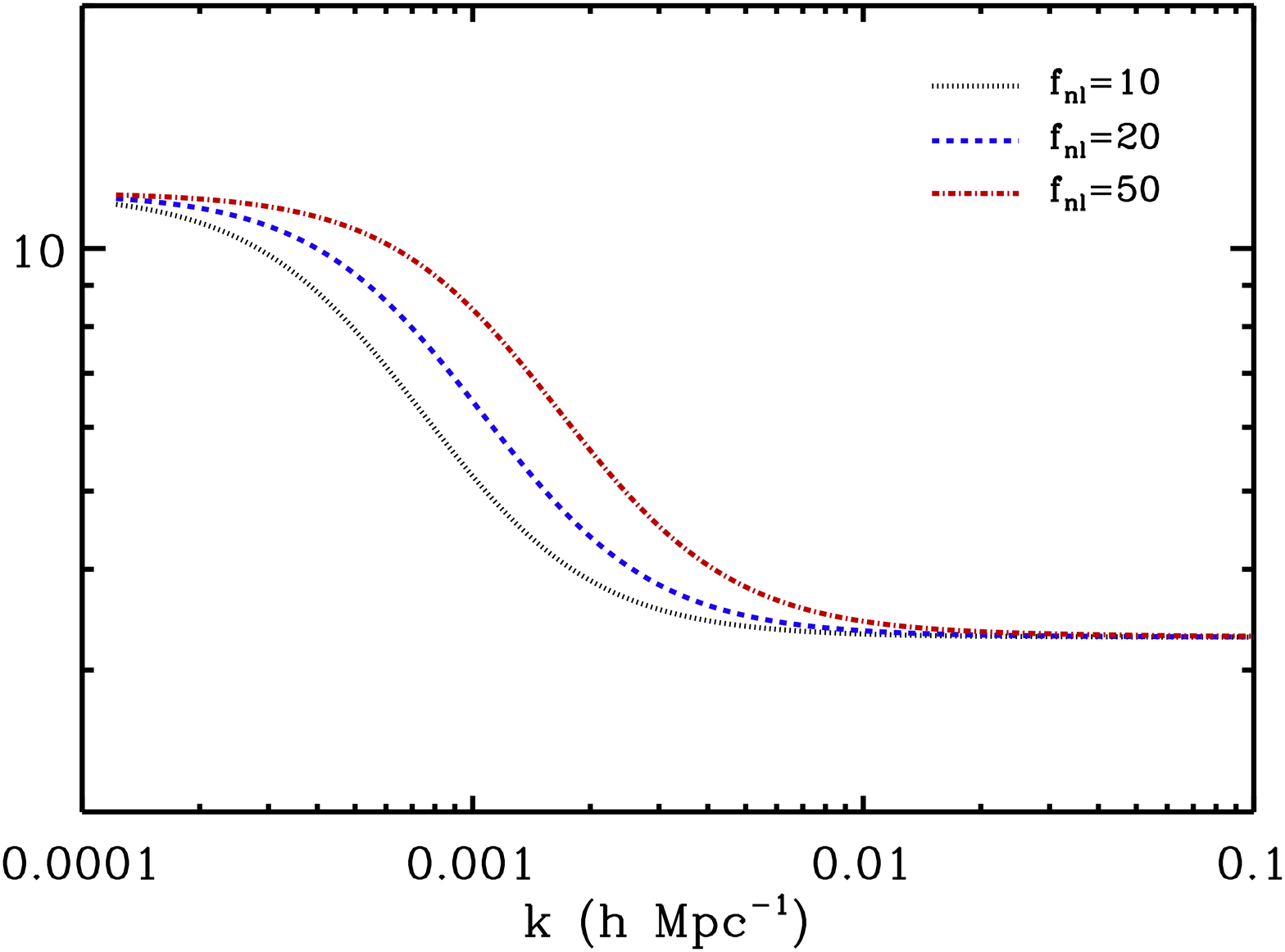}  
  \end{minipage}
  \captionof{figure}{\emph{Left:} 3D halo power spectrum at z$=$1 for different values of $f_{nl}$ with an effective halo bias (eq. \ref{eq:6}) computed in the mass range from $10^{11}h^{-1}M_{\odot}$ to $10^{12}h^{-1}M_{\odot}$(lower curves) and from $10^{13}h^{-1}M_{\odot}$ to $10^{14}h^{-1}M_{\odot}$ (Upper curves). \emph{Right:} Ratio between the $P^{3D}_h$ curves presented on the left panel for the same values of $f_{nl}$.}  
  \label{fig1}
\end{figure*}

As shown in the previous section, the total scale dependent correction introduced by non-Gaussianity in halo clustering ($\delta_h=b_h\delta_m$) depends on the redshift and also on the mass of the halo. In the absence of non-Gaussianity, this mass dependence simply translates into an increased amplitude in the power spectrum or correlation function for more massive halos/galaxies when compared to lower mass objects. 
However, if non-Gaussianity is present, the mass dependence of the linear bias in eq. \ref{eq:2} will introduce different scale dependence features in clustering 2-point statistics. More massive halos will then be more sensitive to non-Gaussianity than lower mass ones. Fig. 1 shows this effect for different values of $f_{nl}$ and halo mass for the 3D halo power spectrum. One can clearly see the differential effect of using mass bins, with the effect being significant only for $k$ less than 0.02 $h$Mpc$^{-1}$, and having a relative amplitude up to order 1 order of magnitude. Such features offers the possibility to obtain additional information on $f_{nl}$ by constructing observational 2-point statistics from objects corresponding to different halo mass ranges. In order to exploit this, we need to address the issue of differentiating halo masses from an observational point of view.

\section{Galaxy bias}

\subsection{Tracers of halo mass}

Most of the studies done to constrain clustering properties of dark matter are based on large surveys of galaxies or clusters that trace the distribution of matter over large scales in the Universe. A key issue when modeling the data obtained from LSS surveys is to know how galaxies populate dark matter halos. This is a complex issue since galaxy formation involves non-linear collapse inside the collapsed halos and there is also the possibility of merging form different sized halos. One way to address this issue is to adopt the Halo model \citep{CooraySeth} which assumes an Halo occupation Distribution function (HoD) to compute the galaxy power spectrum. In this framework, all the information about the mass dependent features is compressed into a few free parameters such as the minimum mass of haloes that can contain galaxies as observed in a survey. The Halo model is a very useful tool to accurately compute the galaxy power spectrum into the non-linear scales, but by incorporating all types of galaxies in the HoD the information on halo mass at large scales is not taken into account.   

Another way to access the halo clustering properties is to consider specific populations of galaxies and clusters which allow the elimination of the one-halo contribution to the power spectrum, and thus more directly trace each object with its underlying dark matter halo. A typical example are the Luminous Red Galaxies (LRGs), a population of old and relaxed galaxies expected to be free from recent merger activity. Such properties justify the use of these objects to reconstruct the halo density field \citep{Reid} and use the resulting power spectrum to constrain cosmological models \citep{Percival} including the presence of non-Gaussianity \citep{Bernardis}. However, even with this analysis, it is not possible to differentiate the mass of the halos associated with LRGs and constraining $f_{nl}$ is done by means of an effective bias, corresponding to a weighted averaged bias over the mass range expected for halos hosting a given galaxy type (e.g. LRGs):            
\begin{equation}
b_{eff}(z)=\frac{\int b_{h}(M,z)\frac{dn}{dz dM}dM}{\int \frac{dn}{dz dM} dM}\;,
\label{eq:6}
\end{equation}
where $dn/(dz dM)$ is the halo mass function.
The above equation has been used to place constraints on non-Gaussian bias using 3D power spectrum from large available datasets and thus reducing the shot noise component at the expense of some loss of information on $f_{nl}$. 

\begin{figure} 
  \hspace{-8pt}
  \includegraphics[width=0.48\textwidth]{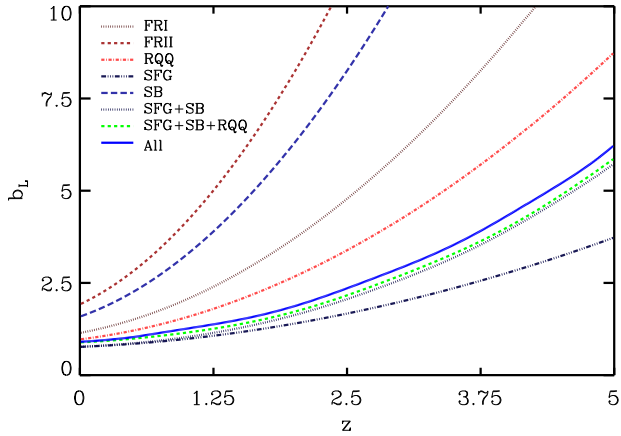}  
  \caption{Bias redshift evolution for the different source combinations considered in this paper, considering a Gaussian distribution of galaxies around the central masses as discussed section \ref{radiopopulations}. When combining populations, the bias is obtained through eq. \ref{eq:11}.}
  \label{fig2}
\end{figure}

\subsection{Radio galaxy populations}
\label{radiopopulations}
A different approach is to consider other specific galaxy populations whose bias properties can be explained by a strong correlation between halo mass and galaxy type.
This seems to be the case for some types of radio galaxies, such as Radio-quiet and Radio-loud AGNs, including FRI and FRII galaxies \citep{FanaroffRiley74}, and also star-forming galaxies, including starbursts. The idea of assigning a single halo mass to these objects was introduced by \cite{Wilman} (W08) in a framework to produce realistic sets of data of the extragalactic sky to be observed with the next generation of radio telescopes.  Indeed, for low redshifts the (scale independent) bias obtained using the formalism presented in section 2.2 with a single halo mass for each population is compatible with the clustering measurements obtained from the NVSS \citep{Condon} and FIRST \citep{Becker} surveys \citep[e.g.][]{BlakeWall2002,Overzier2003,Blake2004,Wilman2003,Lindsay2014}.
The mass to be assigned to each galaxy type is thoroughly discussed in section 2.7 of W08. Based on different observational features and luminosity relations derived from several clustering studies for each population (we report the reader to the references presented in W08) the authors propose to take the following halo masses\footnote{Note that the simulation itself assigned some finite mass interval to each population. This prescription just states that, to a good approximation, the effective bias can be obtained by taking just one halo mass. Moreover, any correction function that translates the number of halos to the actual number of galaxies (as seen in the SKADS simulation), will not affect the bias as long as it is not a function of the mass.}: 
\begin{itemize}
\item Star forming galaxies (SFR): M$_{halo}$=$1\times10^{11}h^{-1}M_{\odot}$ 
\item Starbursts (SB): M$_{halo}$=$5\times10^{13}h^{-1}M_{\odot}$ 
\item Radio Quiet Quasars (RQQ): M$_{halo}$=$3\times10^{12}h^{-1}M_{\odot}$
\item Radio loud AGN (FRI): M$_{halo}$=$1\times10^{13}h^{-1}M_{\odot}$ 
\item Radio loud AGN (FRII): M$_{halo}$=$1\times10^{14}h^{-1}M_{\odot}$
\end{itemize}

In W08, the authors also argued for the introduction of some type of upper limit in the bias of a given population. The argument was that without such limit the bias would become too large at high redshifts. However, there is not enough theoretical evidence supporting such an abrupt cut on the bias at a given redshift. Different numerical simulations \citep{SethTormen,basilakos} suggest that the peak biased formalism of eqs. \ref{eq:2} and \ref{eq:3} is valid at least up to z$\approx 5$ for a wide range of masses. We will therefore in our study consider no restrictions in the bias evolution when computing the angular power spectrum predictions\footnote{This is basically a statement that, the higher the mass (and the rarer) an object is, the larger its bias. The quantity $\frac{dn}{dz} b_h$ (which is what we need for the calculation), on the other hand, does not show this dramatic increase with z.}.

Although this model can describe the observed bias for a significant redshift range, it is obviously subject to a degree of uncertainty. It is obvious that attributing a single halo mass to these large galaxy population types is a very strong assumption and cannot be correct in physical terms. In order to have a more plausible distribution for the halo mass associated with each population, we consider a Gaussian distribution of masses around the central mass values, with a standard deviation of 20$\%$ of the central mass $M_{cent}^i$, which will be considered as a free parameter for each population. The mass values presented above will be considered as fiducial values. In this framework, the linear bias for each population can be reduced to:
\begin{equation}
b^i(z)=\int f^i(M,M_{cent}^i)b_L^i(M,z) dM\;,
\label{eq:12}
\end{equation}
with $f^i(M,M_{cent}^i)$ being the Gaussian distribution of masses mentioned above and $b_L^i(M,z)$ the linear bias from eq. \ref{eq:3}. The choice of function here should in principle depend on the probability of a halo of a given mass to generate a given observed galaxy type as well as on the halo mass function itself. We take the simple approach here of using a Gaussian noting that, as seen later on, the actual shape of the function does not affect the constraints.
Using this model one can study the impact of the mass values in the constraints of $f_{nl}$ via eqs. \ref{eq:12}, \ref{eq:2} and \ref{eq:3}.

\section[]{Forecasts for future experiments}

\subsection{The Fisher matrix}

We discussed above how radio galaxies can trace the mass of their host halos and thus be used to differentiate populations of halos in order to constrain primordial non-Gaussianity models.  
Radio surveys also have the advantage to cover a large fraction of the sky in a relatively small amount of time, with the potential to detect objects up to very high redshifts. However, the sensitivity required to determine redshifts from the 21-cm H{\sc i} emission line makes it difficult to use the data of these surveys to compute the 3D power spectrum on large scales until the full SKA \citep[e.g.][]{Abdalla2010} but see \cite{Camera2013}. Therefore,  we will focus on the angular power spectrum that will be measured by future continuum surveys.

Due to their high sensitivity, experiments conducted by the SKA and its precursors/pathfinders, such as LOFAR and WODAN on APERTIF \citep{Rottgering}, MIGHTEE on MeerKAT \citep{Jarvis2012} and EMU on ASKAP \citep{Norris}, are expected to detect galaxies up to $z \sim 5$ with a large redshift distribution. In this section, we present forecasts on the constraints obtained from the angular auto and cross correlation power spectra that can be measured using samples of different galaxy types.       

The forecasts were performed using the Fisher matrix formalism (\cite{Fisher}, \cite{Tegmark96}). The Fisher information matrix allows the computation of the expected errors on the parameters of a given model assuming that the likelihood function near the best fit model is approximated by a multivariate Gaussian. It also allows us to include different datasets, observables and models in a consistent way.
The full Fisher matrix, when using multi-tracers is then the sum of the Fisher matrices for each multipole: 
\begin{equation}
 F_{\alpha\beta}=\sum_{\ell=\ell_\mathrm{min}}^{\ell_\mathrm{max}}\frac{2\ell+1}{2}f_\mathrm{sky}\text{Tr}\left(\frac{\partial \mathbfss C(\ell)}{\partial\theta_\alpha}\left(\bf{\Gamma}_\ell\right)^{-1}\frac{\partial \mathbfss C(\ell)}{\partial\theta_\beta}\left(\bf{\Gamma}_\ell\right)^{-1}\right) .\label{eq:Fisher}
\end{equation}
In the above expression the matrix $\mathbfss C(\ell)=\left[C^{ij}\right](\ell)$ contains all the auto and cross angular power elements corresponding to tracers $i$ and $j$ and $\theta_{\alpha}$ is a given parameter of the model to be constrained.  
For the fiducial cosmological model, we considered a standard $\Lambda CDM$ obtained from the latest Planck release \citep{Planck1}, SDSS LRG power spectrum and H$_0$ \citep{Komatsu}. Since the only cosmological parameters that can be potentially degenerate with the non-Gaussian bias are the matter density $\Omega_m$, Hubble parameter $H_0$ and primordial fluctuation amplitude $A_S$, these were introduced in the Fisher matrix as free parameters in order to make a consistent analysis, adding at the same time a Fisher matrix corresponding to the covariance matrix for these parameters obtained by the same combination of data mentioned above.

Finally, we added the local non-Gaussianity parameter $f_{nl}$ and the central mass for each galaxy population $M_{cent}$ describing the halo mass distribution. As for the dispersion of these distributions, a fixed value of 0.2 was taken, corresponding to 20$\%$ of the central mass value for each population. The full parameter set used in the Fisher analysis was then $\boldsymbol{\mathcal{P}}=\{\Omega_m; h; \log 10^{10 A_s};f_{nl};M_{cent}^{SFG};M_{cent}^{SB};M_{cent}^{RQQ};M_{cent}^{FRI};M_{cent}^{FRII} \}$ with  $\boldsymbol{\mathcal{P}}_{fiducial}=\{0.308 ; 0.678 ; 3.091 ; 0 ; 10^{11}M_{\odot}/h ; 5\times10^{13}M_{\odot}/h ; 3\times 10^{12}M_{\odot}/h ; 10^{13}M_{\odot}/h ; 10^{14}M_{\odot}/h\}$.

The model presented in Sec. \ref{section2} was applied to compute the $C_{\ell}$ and the covariance matrix $\Gamma$ was considered to follow:
\begin{equation}
\Gamma^{ij}_{\ell}= C^{ij}_{\ell} + \delta^{ij} N^{ii}\;.
\label{eq:10}
\end{equation}
Here $N^{i}$ is the noise power spectrum for a given galaxy population which is determined by $1/n^{i}$, and $n^{i}$ is the number of sources per steradian in the sky. This last expression also assumes that all the systematics inherent to each survey are sub-dominant to the cosmic variance and shot noise contributions.

\subsection[]{The multi tracer technique in the Fisher matrix formalism}

In the Fisher matrix formalism, the constraining ability of an experiment for a given parameter is determined by the model dependence on the parameter and the covariance matrix of the observational data. More precisely, if the inverse of the covariance matrix has sufficiently large diagonal elements, the precision that can be attained on a parameter can reach sub-percentage levels. The Fisher matrix used above is derived assuming that the measurements are the maps themselves, e.g. the likelihood function is the probability distribution of the $a_{lm}$, the spherical harmonic transform of the galaxy number densities observed on the sky. Since we are comparing maps with different bias, the possible cancellation of cosmic variance in the bias measurement due to the use of multiple tracers is already implicitly included in the formalism. 
As an illustration, one can consider a hypothetical ``perfect'' experiment measuring two galaxy populations with enough sensitivity so that the shot noise contribution to eq. \ref{eq:10} can take any small value. In this case the covariance matrix is given by: 

\begin{figure}
  \hspace{-8pt}
     \includegraphics[width=0.495\textwidth]{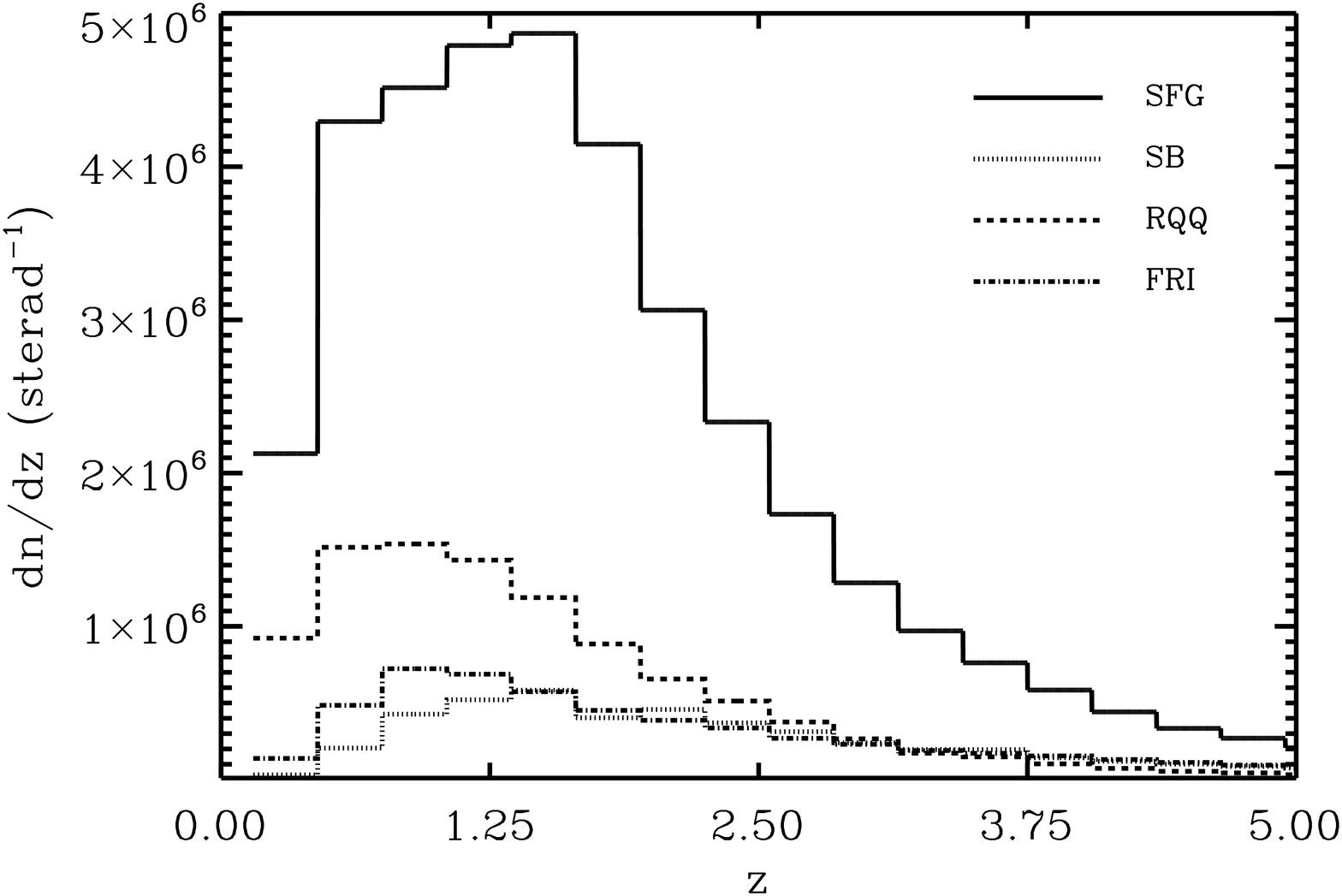}  
   \caption{Redshift distribution of sources per steradian for SKA phase-1 (Flux cut detection at 5 $\mu$Jy). The source types are Star Forming Galaxies (SFG), Starburst galaxies (SB), Radio quiet quasars (RQQ) and FRI. We omit the distribution for FRII galaxies since their number is much lower (of the order of 100) and would not be visible in this figure. All distributions were obtained form the S$^3$ catalogs down to a sensitivity limit of 1$\mu$Jy and applying a cut at 5$\sigma$.}
  \label{fig3}
\end{figure}

\begin{equation}
\bf{\Gamma} \propto
 \begin{bmatrix}

       \int k^2 P_\delta(k) W_1^2(k) dk+\frac{1}{n_1}  &   \int k^2 P_\delta(k) W_1(k)W_2(k) dk \\[0.3em]
       \int k^2 P_\delta(k) W_1(k)W_2(k) dk &   \int k^2 P_\delta(k) W_2^2(k) dk+\frac{1}{n_2}  \\[0.3em]
     \end{bmatrix}
\label{matrix_cov2}
\end{equation}

 If population 2 presents a bias and redshift distribution so that the product $b_2 dn_2/dz$ is proportional by a factor of A to that of population 1 ($b_2 dn_2/dz=A b_1 dn_1/dz$), the covariance matrix above is reduced to the simple form:

\begin{equation}
\bf{\Gamma} \propto
 \begin{bmatrix}
     
       1+\frac{1}{n_1} & A   \\[0.3em]
        A & A^2+\frac{1}{n_2} .\\
     \end{bmatrix}
 \label{matrixsimp}
\end{equation}

\begin{figure*}
   \hspace{0pt}
  \centering
  \begin{minipage}[b]{8.6cm}
    \includegraphics[width=\textwidth]{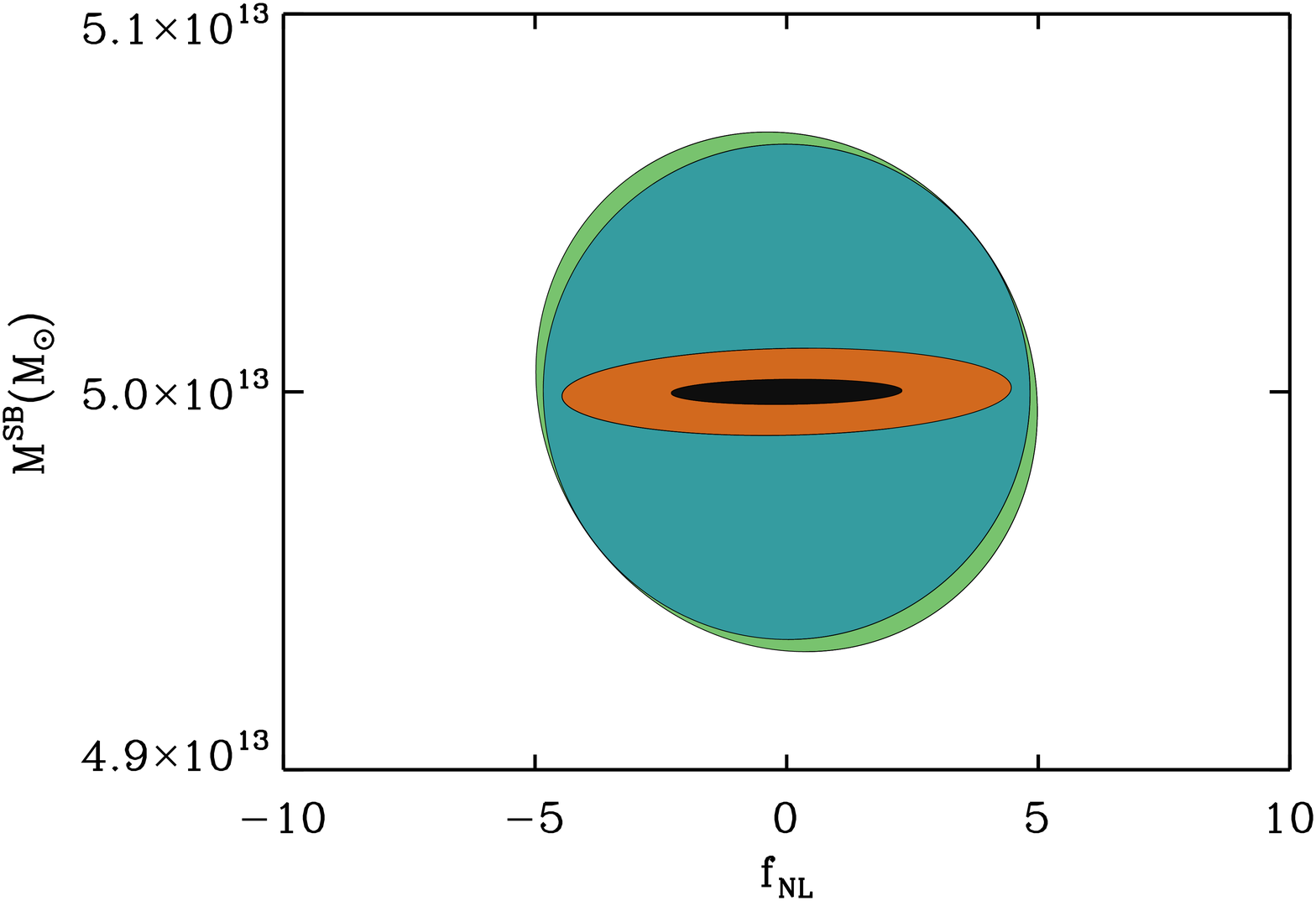}  
  \end{minipage}
  \begin{minipage}[b]{8.6cm}
    \includegraphics[width=\textwidth]{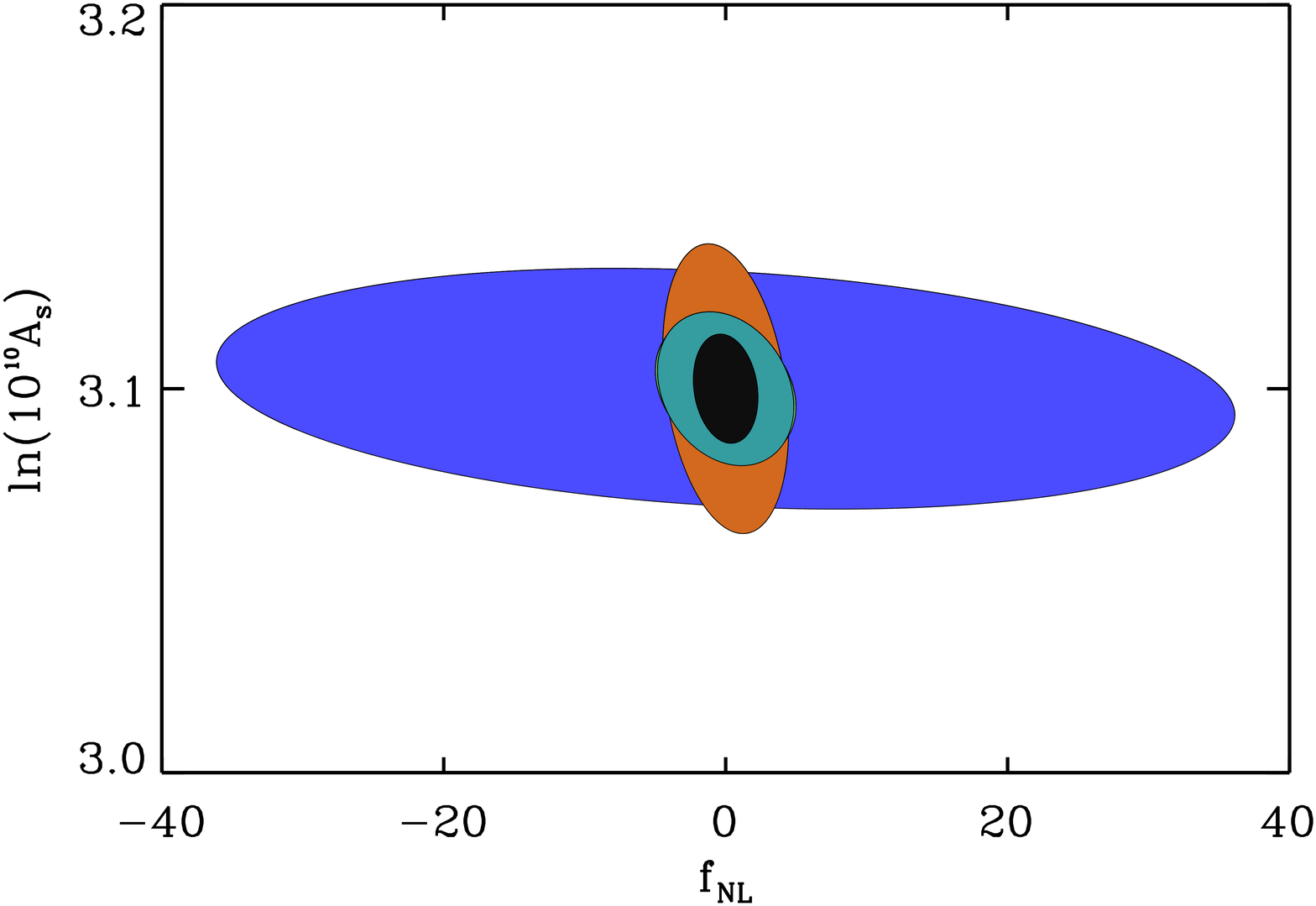}  
  \end{minipage}
  \caption{1-$\sigma$ forecasts on $f_{nl}$ and central mass for starburst population (\emph{left}) and on $f_{nl}$ and primordial fluctuation amplitude (\emph{right}), both for a flux cut detection at 5 $\mu$Jy. Larger blue ellipse considers the whole sample of galaxies and light green ellipses represent the constrains if we consider the combination of SFG, SB and RQQ (or just SFG and SB, in light blue) as one single undifferentiated population with bias defined by eq. \ref{eq:11}. The orange contour corresponds to the constraint using the whole differentiated 5 galaxy population for z$<$1 and 4  bins with SFG and SB undifferentiated for z$>$1. Finally, the smaller and darker ellipse shows the ideal case where all 5 populations could be differentiated over the entire redshift range.}
  \label{fig4}
\end{figure*}
If such approximation was correct this would mean that as one approaches a perfect measurement with no noise, the inverse of the covariance matrix would go to infinity and one would have a nearly perfect estimation of $f_{nl}$. We illustrate this effect considering a case with $A=4$, and $b_1=b_h(z)$ for a halo mass of $10^{11} h^{-1}M_{\odot}$ given by eqs. \ref{eq:2} and \ref{eq:3}. We used a generic wide redshift distribution extending up to z=6 with a peak at z=1. The Fisher matrix was computed with a numerical code in the following possibilities: 1) using population 1 alone, which provides a constraint of $\sigma_{f_{nl}}=50$. ii) using population 2 alone, with its higher window function value, which yields a better constraint of $\sigma_{f_{nl}}=14$. iii) using information from both populations as in eq. \ref{matrix_cov2}, where the improvement is dramatic, with a constraint of $\sigma_{f_{nl}}=2\times10^{-13}$. By including different biased tracers in the same data analysis one can indeed significantly improve the constraints on $f_{nl}$ and avoid the cosmic variance limitation. This happens because we are observing the same sky for both populations and the main difference in the clustering of dark matter and galaxies will be the non-Gaussianity induced boost at large scales. This effect was first introduced by \citet{Seljak} and further explored in \citet{Abramo}, both focusing on the 3D power spectra for galaxy redshift surveys.   
 The example presented above is somehow pedagogical, and in reality the window function for two different populations will not be proportional and hence the cosmic variance cancellation effect will be somehow diminished. In the following section we focus on quantifying the constraints that can be put on $f_{nl}$ by future experiments using realistic modeling for both bias and galaxy distributions.

\subsection{Requirements of radio surveys for galaxy differentiation}
\label{requirements}

The key to use the multi-tracer analysis presented in this paper is the ability to identify the galaxy populations using observational features in a given radio galaxy survey. Starting with radio-loud AGN, these can be divided into FRI or FRII radio galaxies, which can be distinguished by the presence of coaxial lobes of emission which are offset from the core and present emission hot spots in the lobes edges (FRII) and those where the lobes are connected to the core by jets and present a more diffuse surface brightness (FRI) \citep{Gendre2010,Gendre2013}. Such distinction can be made only with a sufficient angular resolution experiment that allows to identify these morphological differences. Following \cite{Wilman}, the typical size of the gap between the lobes is modeled as $0.2(1+z)^{-1.4}$Mpc. This implies an angular resolution of about 1 arcsec at worst to distinguish FRI and FRII galaxies at high redshifts such as z=4. This requirement means that such a study is possibly only feasible with Phase I of the SKA itself, given that the various precursor and pathfinder surveys all have substantially poorer resolution than this (i.e. $\sim 5$~arcsec in the case of MeerKAT and 10-15~arcsec for EMU and WODAN, although LOFAR could potentially reach 1~arcsec resolution with inclusion of the longer baselines). Phase I of the SKA will also have extremely high sensitivity to diffuse emission due to the envisaged compact core, thus providing the necessary baselines to probe all spatial scales associated with AGN in a single survey.
As for star-forming galaxies (SFGs) and starbursts, it is be extremely difficult to distinguish them with radio data alone, as the only distinction is in their star formation rates which are difficult to access without redshift information for the individual sources.

Other sensitive issue is the distinction of radio quiet AGNs and SFG/SB galaxies. Multi-wavelength studies on different deep fields suggest that radio quiet quasars may account for almost half of the sources observed at a level of a few $\mu$Jy \citep[e.g.][]{Simpson2012,McAlpine2013}. The exact identification of this source type in a large continuum survey may require a multi-wavelength analysis, in particular, future X-ray surveys such as eROSITA \citep[e.g.][]{Kolodzig2013}, since most quasars emit a large fraction of the radiation in the X-ray band, as opposed to star forming galaxies. To overcome this issue, and since the goal of the present paper is to lay the foundations of this multi-tracer analysis method, we will consider and present forecasts for the cases where RQQ, SFG and SB could be successfully distinguished and also where they have to be merged into different population combinations.
For this case, the total bias of a merged sample becomes (from eq. \ref{eq:6} and \ref{eq:12}): 
\begin{equation}
b^{tot}(z)=\frac{\sum_i n^i(z)b^i(z)}{\sum n^i(z)}\;,
\label{eq:11}
\end{equation}
where $n^i(z)$ is the total number of galaxies (or number density) expected at a given redshift and $b^i(z)$ is defined in equation \ref{eq:12}.
The redshift evolution for the bias of each population, and for different merged sub-populations is presented in fig. \ref{fig2}.  
  
  \begin{table*}
\centering
\begin{tabular}{cccccc}
\hline
  &  \multicolumn{4}{c}{$\sigma_{f_{nl}}$} \\
\hline
Flux detection threshold  &  Full sample$^a$ & 3 bins & 4  & 4+5 bins & 5 bins \\
1 $\mu$Jy & $ 12\;\;(9.6) $ & $ 2.8 $ & $ 2.7 $ & $ 2.2 $ & $ 0.7 $ \\
3 $\mu$Jy & $ 25\;\;(17) $ & $ 2.7 $ & $ 2.7 $ & $ 2.6 $ & $ 1.2 $ \\
5 $\mu$Jy & $ 32\;\;(23) $ & $ 3.3 $ & $ 3.2 $ & $ 2.9 $  & $ 1.5$ \\
10 $\mu$Jy & $ 48\;\;(35) $ & $ 3.7 $ & $ 3.7 $ & $ 3.6 $ & $ 1.9 $ \\
\hline
\end{tabular}
\caption{Forecasts on $f_{nl}$ 1-$\sigma$ errors using the angular power spectra of different galaxy populations for different detection flux limits. We present the results obtained using the full sample of objects with an averaged effective bias and those obtained using the combination of 3 populations of radio galaxies (where SRG, SB and RQQ correspond to one population group), using 4 populations (where only SFG and SB are undifferentiated) and with a selection of 5 populations for z$<$1 and 4 populations for z$>1$ (again with undifferentiated SFG and SB). We also show the result for the ideal case where all 5 populations could be differentiated over the entire redshift range of the survey.\newline
$^a$ Values in parentheses were obtained assuming a constant and single mass for each sub-population.}
\label{tab1}
\end{table*}

\begin{table*}
\centering
\begin{tabular}{cccc}
\hline
  &  \multicolumn{2}{c}{$\sigma_{f_{nl}}$} \\
\hline
Change in fiducial mass & 3 bins & 4 bins & 5 bins \\
\hline
\hline
SB -20\% & $ 3.2 $ & $ 3.2 $ & $ 1.6 $ \\
SB +20\% & $ 3.3 $ & $ 3.2 $ & $ 1.4 $  \\
FRI -20\% & $ 3.5 $ & $ 3.5 $ & $ 1.5 $ \\
FRI +20\% & $ 3.1 $ & $ 3.0 $ & $ 1.5 $  \\
SB -20\%, FRI -20\% & $ 3.6 $ & $ 3.4 $ & $ 1.6 $  \\
SB -20\%, FRI +20\% & $ 3.1 $ & $ 2.9 $ & $ 1.6 $  \\
SB +20\%, FRI -20\% & $ 3.6 $ & $ 3.5 $ & $ 1.5 $  \\
SB +20\%, FRI +20\% & $ 3.1  $ & $ 3.1 $ & $ 1.5 $  \\
\hline
\end{tabular}
\caption{Same as Table \ref{tab1}, but assuming a 20\% variation on the fiducial central mass for Starburst and FRI galaxy halos.}
\label{tab2}
\end{table*}           

\subsection{Surveys with SKA-phase 1}

Phase~1 of the SKA has the potential to carry out a 3$\pi$~sr radio continuum survey down to 1$\mu$Jy rms with sub-arcsecond resolution and, as described in the previous section, with a large number of short baselines for enhanced sensitivity to diffuse emission. Such a survey, given the current specifications would require about 2~years on sky and would be valuable for many areas of astronomy and cosmology \citep[e.g.][]{Raccanelli,Camera2012,Rubart2013}.

We used the S$^3$ simulation database \citep{Wilman} to obtain the redshift distribution of each source corresponding to surveys with flux-density limits to encompass any future surveys with the SKA or similar. More precisely, we considered flux-density cuts of 1, 3, 5 and 10$\mu$Jy (the rms noise should be about 5 times smaller). In particular, the total redshift distribution of sources is shown to have a median redshift of 1.1, with a total number of detected galaxies and quasars of $\sim 5\times10^8$ for SKA-phase 1 assuming a 5$\sigma$ flux density limit of $5\mu$Jy. The number distribution of sources for each population considered in this paper is presented in fig. \ref{fig3}. We used these to determine the shot noise contribution to the covariance matrix for each population and fitted them in order to obtain the $dn/dz$ used to compute both  the Fisher and covariance matrices. 

\subsection{Results}

Given the large sky coverage of the considered surveys, we considered $\ell_{min}=2$ and $\ell_{max}=200$ to compute the Fisher matrix, assuming that these scales contain most of the information on non-Gaussianity.
The results of the forecasts on $f_{nl}$ for different flux detection limits are presented in table \ref{tab1}. The improvement of these constraints when using all the galaxy multi-tracer information compared to the undifferentiated whole galaxy sample is remarkable. For example, for a 5$\mu$Jy cut, corresponding to the expected detection limit of SKA phase 1,  we obtain $\sigma_{f_{nl}}\approx 23$ for the latter and $\sigma_{f_{nl}}\approx 1.5$ for the former using 5 multi-tracer galaxy populations. However, since the differentiation of Starburst and SFG galaxies is problematic (as discussed in Sec.\ref{requirements}), a more realistic case would consider 5 distinguishable populations up to z$<$1 (where SB and SFG's would be distinguished using a deep optical survey) and a merged populations of SB and SFG for $z>1$. For this case the improvement is still quite significant, with $\sigma_{f_{nl}}\approx 2.9$.
We also checked for this particular case that imposing $l_{max}=400$ does not affect the result, which takes exactly the same value.

\begin{figure}
  \hspace{-5pt}
  \centering
    \includegraphics[width=0.48\textwidth]{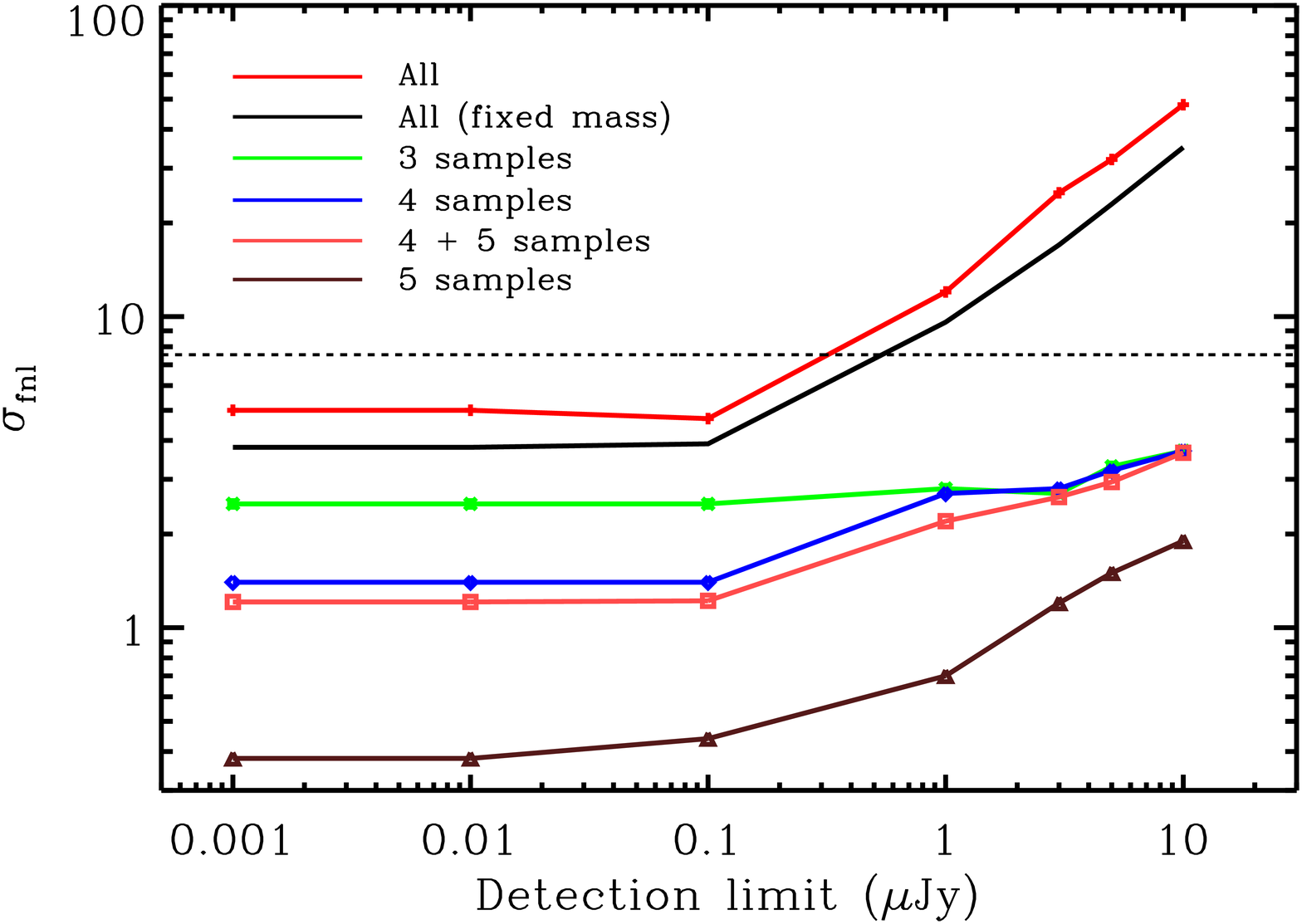}  
  \caption{Forecasted constraints on $f_{nl}$ obtained with the multi-tracer method as a function of the flux cut used to detect galaxies. The horizontal line represents the best constrain obtained by the Planck collaboration \citep{Planck2} normalized to the convention we are using here.}
  \label{fig5}
\end{figure}

The impact of allowing the central mass for each population as a free parameter is somehow small, except when combining all the populations, because this introduces a large uncertainty on the bias evolution. To limit such uncertainty, we considered two other ways to compute the bias for this case (where all populations are combined): fixing the mass of each population to their fiducial values and thus taking eqs. \ref{eq:12} and \ref{eq:11} with $f^i(M,M_{cent})=\delta(M-M_{cent})$ and also using eq. \ref{eq:6} within the mass range of $M_{min}=1\times10^{11} h^{-1}M_{\odot}$ to $M_{max}=1\times10^{14} h^{-1}M_{\odot}$(equivalent to the usual effective bias based on the halo model). Both of these approaches yield almost the same results, and improve the constraints on $f_{nl}$ by around 25\%. (see table \ref{tab2} and fig. \ref{fig5})     
Another important result is the fact that the variance assumed for the Gaussian distributions of the halo masses for each population does not significantly effect the results for the multi-tracer analysis. We tested using smaller values such as 1 \% and larger values such as 40 \% and the differences obtained in $f_{nl}$ and M$_{cent}$ did not exceed 2\%.

Note that the degeneracy between the non-Gaussianity parameter and central mass is almost absent, as shown in fig. \ref{fig4}). The cosmological parameters can show a more significant correlation with $f_{nl}$, such as is the case of the primordial fluctuations amplitude $A_S$.

It is also important to note that some populations contribute more to the constraint on $f_{nl}$ than others. The overall contribution from a single galaxy population depends mainly on the combination of their bias and measured shot noise, and to lesser extent, on their redshift distribution. We noted in our simulation that the Starburst and FRI galaxies had the most important contribution. Although the inclusion of M$_{cent}$ as a free parameter is important to access the impact of the uncertainty in these values, the fiducial value we take to perform the constraints can also have a significant importance. We tested this considering both populations that contribute more to the constraints and varying their central mass values by 20\% (both larger and smaller). The results on $\sigma_{f_{nl}}$ are sumarized in table \ref{tab2}. FRI fiducial masses seem to contribute the most to the results, but overall the constraints are stable up to 17 \%.  

Finally, we tested what would happen if somehow one could have even better detection sensitivities. This was done with the purpose of identifying the absolute limit in precision that can be achieved with the multi-tracer technique. In fig. \ref{fig5} we present the results for all the considered data combinations, where one can see that for most cases, the 1$\mu$Jy detection limit represents a sort of a plateau, with the best possible result (again, in the more difficult observational configuration) being an impressive $\sigma_{f_{nl}}\approx 0.4$ at 0.001$\mu$Jy.

\section[]{Summary and Conclusions}

In this paper we studied the impact  of using future radio surveys to constrain local primordial non-Gaussianity which is connected to fundamental physics during the period of inflation. Specifically, we investigated the potential of using observational features of galaxies to differentiate them in terms of the mass of their host halo and thus allowing us to use multi-tracer objects with distinct bias properties to improve existing constraints.

Using simulated catalogs, we show that by taking into account all the statistical information from continuum galaxy surveys, namely the combined auto and cross correlation angular power spectra of multi-tracers galaxies, it is possible to significantly reduce the impact of cosmic variance at large scales, where the effect of non-Gaussianity is expected to be more important. Using the Fisher matrix formalism, we forecast that this method can constrain the $f_{nl}$ parameter up to an accuracy of $\sigma_{f_{nl}}\sim 0.7$, in the optimal case, and $\sigma_{f_{nl}}\sim 2.9$ in a more realistic framework (taking into account the specifications for SKA1). The improvement obtained by considering this multi-tracer analysis is indeed significant, as we have shown that by using the whole galaxy catalog without mass (galaxy type) differentiation only allows one to constrain $f_{nl}$ with an error of $\sigma_{f_{nl}}\sim 32$ in the same realistic framework.

Moreover, we tested and showed that these results are robust even if we introduce more degrees of freedom in the linear bias model through the halo average masses for each populations, which are rather uncertain at high redshifts, although based on observational data.
Even if we consider a significant change in some fiducial values for the halo mass of each galaxy type, the multi-tracer analysis still presents the same order of improvement on non-Gaussianity constraints, with differences that can be considered as systematics of order of 10 to 15 \%.
One should note however that the practical application of such method is prone to other systematic uncertainties, the main one being the capability to correctly identify each galaxy type. The results obtained in the present paper show that radio surveys can reach and possibly surpass the level of precision obtained with CMB data and approach the point where non-Gaussianity can be ruled out with more confidence or even be detected at level expected for most inflationary models.

\section*{Acknowledgments}

LF and MGS acknowledge financial support from Portuguese Funda\c{c}\~ao para a Ci\^encia e Tecnologia (FCT-Portugal) under projects PTDC/FIS/100170/2008 and PTDC/FIS-AST/2194/2012. SC is funded by FCT-Portugal under Post-Doctoral grant SFRH/BPD/80274/2011. MJJ acknowledges support from the South African Square Kilometer Array Project and the South African National Research Foundation.

\appendix

\bsp

\label{lastpage}

\end{document}